# Title: Constructing oxide interfaces and heterostructures by atomic layer-by-layer laser molecular beam epitaxy

**Authors:** Qingyu Lei[1], Maryam Golalikhani[1], Bruce A. Davidson[1,2], Guozhen Liu[1], D. G. Schlom[3,4], Qiao Qiao[1,5], Yimei Zhu[5], Ravini U. Chandrasena[1], Weibing Yang[1], Alexander X. Gray[1], Elke Arenholz[6], Andrew K. Farrar[7], Dmitri A. Tenne[7], Minhui Hu[8], Jiandong Guo[8], Rakesh K. Singh[9], X. X. Xi[1]*

**Affiliations:**

[1]Department of Physics, Temple University, Philadelphia, PA 19122, USA.

[2]CNR-Istituto Officina dei Materiali, TASC National Laboratory, Trieste, I-34146, Italy.

[3]Department of Materials Science and Engineering, Cornell University, Ithaca, NY 14853, USA.

[4]Kavli Institute at Cornell for Nanoscale Science, Ithaca, New York 14853, USA.

[5]Department of Condensed Matter Physics and Materials Science, Brookhaven National Laboratory, Long Island, NY 11973, USA.

[6]Advanced Light Source, Lawrence Berkeley National Laboratory, Berkeley, CA 94720, USA

[7]Department of Physics, Boise State University, Boise, ID 83725, USA.

[8]Beijing National Laboratory for Condensed Matter Physics and Institute of Physics, Chinese Academy of Sciences, Beijing, 100190, P. R. China.

[9]School of Materials, Arizona State University, Tempe, AZ 85287, USA.

*Correspondence to: e-mail: xiaoxing@temple.edu

**Abstract**: Advancements in nanoscale engineering of oxide interfaces and heterostructures have led to discoveries of emergent phenomena and new artificial materials. Combining the strengths of reactive molecular-beam epitaxy and pulsed-laser deposition, we show here, with examples of $Sr_{1+x}Ti_{1-x}O_{3+\delta}$, Ruddlesden-Popper phase $La_{n+1}Ni_nO_{3n+1}$ ($n = 4$), and $LaAl_{1+y}O_{3(1+0.5y)}$/$SrTiO_3$ interfaces, that atomic layer-by-layer laser molecular-beam epitaxy (ALL-Laser MBE) significantly advances the state of the art in constructing oxide materials with atomic layer precision and control over stoichiometry. With ALL-Laser MBE we have produced conducting $LaAlO_3$/$SrTiO_3$ interfaces at high oxygen pressures that show no evidence of oxygen vacancies, a capability not accessible by existing techniques. The carrier density of the interfacial two-dimensional electron gas thus obtained agrees quantitatively with the electronic reconstruction mechanism.

**Main Text:**

Technological advances in atomic-layer control during oxide film growth have enabled the discoveries of new phenomena and new functional materials, such as the two-dimensional (2D) electron gas at the $LaAlO_3/SrTiO_3$ interface (*1, 2*) and asymmetric three-component ferroelectric superlattices (*3, 4*). Reactive molecular-beam epitaxy (MBE) and pulsed-laser deposition (PLD) are the two most successful growth techniques for epitaxial heterostructures of complex oxides. PLD possesses experimental simplicity, low cost, and versatility in the materials to be deposited (*5*). Reactive MBE employing alternately-shuttered elemental sources (atomic layer-by-layer MBE, or ALL-MBE) can control the cation stoichiometry precisely, thus producing oxide thin films of exceptional quality (*6-8*). There are, however, limitations in both techniques. Reactive MBE can use only source elements whose vapor pressure is sufficiently high, excluding a large fraction of $4d$ and $5d$ metals. In addition, ozone is needed to create a highly oxidizing environment while maintaining low-pressure MBE conditions, which increases the system complexity. On the other hand, conventional PLD using a compound target often results in cation off-stoichiometry in the films (*9, 10*). In this paper we present an approach that combines the strengths of reactive MBE and PLD: atomic layer-by-layer laser MBE (ALL-Laser MBE) using separate oxide targets. Ablating alternately the targets of constituent oxides, for example SrO and $TiO_2$, a $SrTiO_3$ film can be grown one atomic layer at a time. Stoichiometry for both the cations and oxygen in the oxide films can be controlled. Although the idea of depositing atomic layers by PLD has been explored since the early days of laser MBE (*11, 12*), we show that levels of stoichiometry control and crystalline perfection rivaling those of reactive MBE can be achieved by ALL-Laser MBE. The technique is effective for both non-polar (such as SrTiO$_3$) and polar materials, such as the Ruddlesden–Popper (RP) phase La$_{n+1}$Ni$_n$O$_{3n+1}$ with $n = 4$. By growing LaAlO$_3$ films on SrTiO$_3$ substrates at an oxygen pressure of 37 mTorr, sufficiently high to alleviate oxygen deficiency in SrTiO$_3$, we show that the properties of the 2D electron gas at the LaAlO$_3$/SrTiO$_3$ interface are in quantitative agreement with the electronic reconstruction mechanism.

The principle of ALL-Laser MBE is schematically illustrated in Fig. 1A. The key difference between ALL-Laser MBE and conventional PLD or laser MBE is the use of separate oxide targets – instead of using a compound target of SrTiO$_3$, targets of SrO and TiO$_2$ are switched back and forth as they are alternately ablated by a UV laser beam. In conventional PLD or laser MBE using a compound target, all elements are ablated at once and the film grows unit cell by unit cell. In ALL-Laser MBE using separate targets, on the other hand, the film is constructed one atomic layer at a time. The number of laser pulses on each target for one atomic layer is around 100, allowing a stoichiometry control of about 1%.

It has been a common practice to control the layer-by-layer growth of thin film by recording and analyzing in real time its reflection high-energy electron diffraction (RHEED) pattern (*13, 14*). The intensity of the specularly-reflected RHEED spot is commonly used, which oscillates depending on the step edge density of the film. One oscillation period corresponds to the deposition of one unit cell layer in the unit cell-by-unit cell growth (*15*). Haeni *et al.* have found that the intensity of the diffracted spot can be used to control the growth of each atomic layer of SrTiO$_3$ films in reactive MBE with alternately shuttered growth (*7*). In this work, we also use the diffracted spot intensity oscillation and our results confirm that the phenomenology identified by Haeni *et al.* also applies to ALL-Laser MBE. Figure 1B shows the RHEED intensity oscillations as the targets of SrO and $TiO_2$ are alternately ablated. Starting from a $TiO_2$–terminated $SrTiO_3$ substrate surface, the diffracted spot intensity increases to a maximum when one monolayer of SrO is deposited; it then decreases to a minimum when one monolayer of $TiO_2$ is subsequently deposited. Furthermore, we have found that the specular spot also oscillates with the same period as the diffraction spot, albeit 180° out of phase, if the Kikuchi lines caused by the diffused scattering of electrons do not overlap the specular spot (see Supplementary Materials).

The RHEED intensity depends on both the surface step edge density and the surface chemistry. When all elements of the film are delivered at the same time in the unit cell-by-unit cell growth, the chemistry information is averaged out and only the step edge density of the film is reflected in the RHEED intensity. For ALL-MBE or ALL-Laser MBE, the surface chemistry changes when different atomic layers are deposited sequentially; consequently both the step edge density and chemistry information can be observed. A detail discussion can be found in the Supplementary Materials.

In our experiment, the RHEED diffracted intensity oscillations along the $SrTiO_3$ [110] azimuth were used to calibrate and control the film growth. As shown in Fig. 1B, Sr/Ti > 1 leads to an increasing peak intensity and the appearance of a "double" peak, while Sr/Ti < 1 leads to a reduced peak intensity; Sr/Ti = 1 results in oscillation peaks with a constant intensity and shape. Furthermore, insufficient or excess pulses in each cycle cause beating of the RHEED intensity (Fig. 1C) while the intensity remains constant for 100% layer coverage (Fig. 1D). Using the RHEED intensity oscillation combined with the calibration of laser pulses per atomic layer obtained from the film thickness measurement, the cation stoichiometry in the films can be controlled to within ±1%.

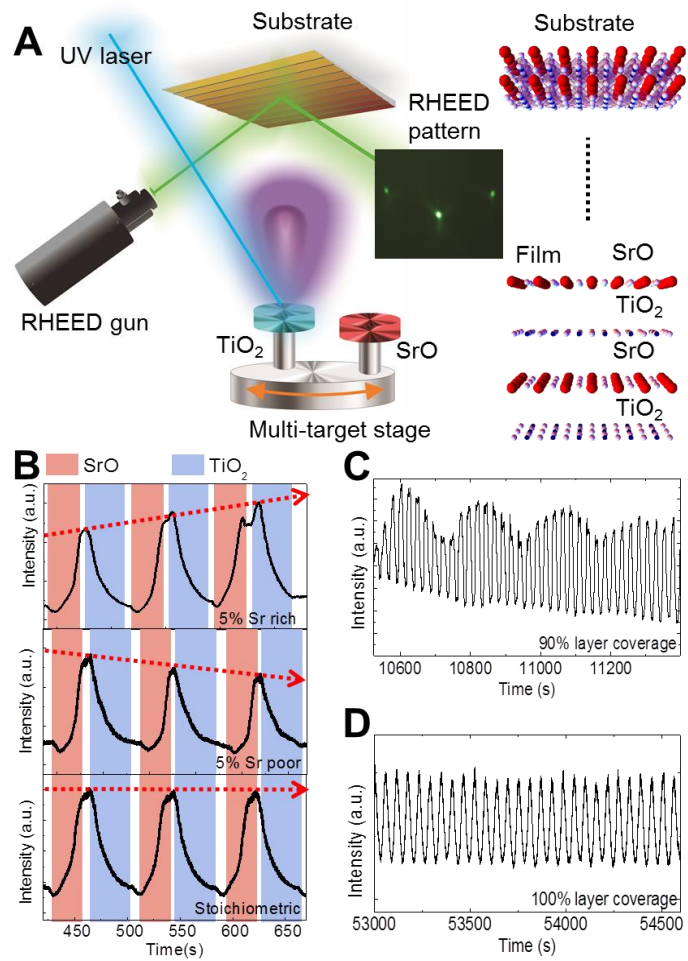

**Fig. 1.** (**A**) Schematic of the ALL-Laser MBE setup. SrO and TiO$_2$ targets are alternately ablated by a UV laser beam while the RHEED pattern is being recorded and analyzed in real-time. The SrTiO$_3$ film is constructed in an atomic layer-by-layer manner. (**B**) RHEED diffracted spot intensity oscillations during the atomic layer-by-layer growth of SrTiO$_3$ films. The red and blue shaded areas represent the depositions of SrO and TiO$_2$ layers, respectively. The white areas represent target switching when no ablation takes place. The red dotted lines indicate variations of the intensity for the cases of Sr-rich, Sr-poor, and stoichiometric deposition, respectively. (**C**) RHEED intensity beating when 0.9 monolayers of SrO and TiO$_2$ are deposited during each target switching cycle. (**D**) Characteristic RHEED intensity oscillations during the growth of a stoichiometric SrTiO$_3$ sample with full layer coverage.

The results of a series of 60 nm thick $Sr_{1+x}Ti_{1-x}O_{3+\delta}$ films grown on $TiO_2$-terminated (001) $SrTiO_3$ substrates by ALL-Laser MBE are presented in Fig. 2. The $x$ values for the five films, determined by the Rutherford backscattering spectrometry (RBS) measurement, shown in Fig. 2A, are 0.13, 0.05, -0.01, -0.07, and -0.12, respectively, with a measurement error of ±5%. They are consistent with the intended compositions ($x$ = 0.10, 0.05, 0.00, -0.05, -0.10) controlled by the respective numbers of laser pulses. Figure 2B shows x-ray diffraction (XRD) $\theta$-$2\theta$ scans for the films around the $SrTiO_3$ 002 diffraction peak along with that of the $SrTiO_3$ substrate. When the film is stoichiometric, the XRD spectrum cannot be distinguished from that of the single crystal $SrTiO_3$ substrate. When the film is not stoichiometric, regardless of Sr rich or deficient, a diffraction peak from the film at a smaller angle than the substrate peak is seen, indicating a $c$-axis lattice expansion. The $c$ lattice constant vs. $x$ is plotted for the films in Fig. 2C. Also plotted are data from films grown by reactive MBE for comparison. The results from the two techniques are in agreement with each other. In Fig. 2D, ultraviolet Raman spectra are presented for the five films as well as a stoichiometric film grown by reactive MBE (*16*) and a single crystal $SrTiO_3$ substrate. For bulk $SrTiO_3$, the Raman spectrum shows only the second-order features (*17*), while the spectra of all the nonstoichiometric samples contain strong first-order Raman peaks, indicating the breakdown of the central inversion symmetry. The stoichiometric film shows spectra similar to those of the stoichiometric film grown by reactive MBE (broad and weak second-order Raman peaks). The results demonstrate that ALL-Laser MBE possesses the same excellent stoichiometry control as ALL-MBE.

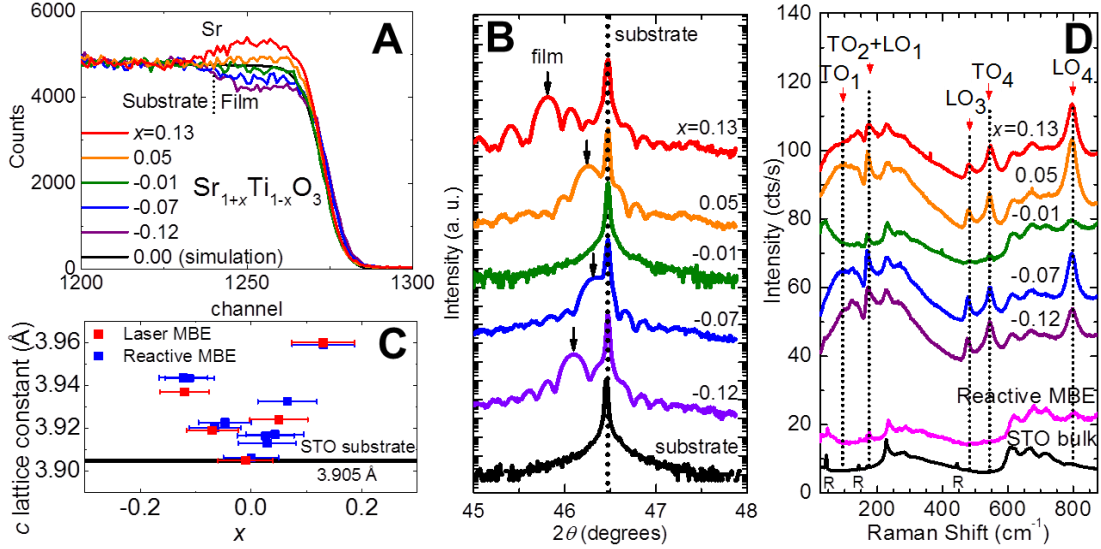

**Fig. 2.** (**A**) RBS spectra for five 60 nm thick $Sr_{1+x}Ti_{1-x}O_{3+\delta}$ films grown by ALL-Laser MBE on $SrTiO_3$ substrates. The compositions determined by RBS have an error of ± 5%. (**B**) XRD $\theta$-$2\theta$ scans of the films and a bare $SrTiO_3$ substrate. The black dotted line represents the substrate peak position, and arrows point to film peaks. (**C**) The $c$ lattice constant (determined by the XRD measurement) vs. $x$ (determined by the RBS measurement) for the five films. The data from films grown by reactive MBE are included for comparison. (**D**) Ultraviolet Raman spectra ($\lambda$ = 325 nm, $T$ = 10 K) for the five films, a stoichiometric film grown by reactive MBE, and a $SrTiO_3$ single crystal.

The ability of ALL-Laser MBE to fabricate layered oxide materials not possible for other techniques is demonstrated with the growth of Ruddlesden-Popper phase $La_{n+1}Ni_nO_{3n+1}$ with $n$=4 on $LaAlO_3$ substrate. Figure 3A shows a schematic of the $La_5Ni_4O_{13}$ structure, in which 4 layers of $NiO_6$ octahedra are sandwiched between the NaCl-type LaO double layers. No successful deposition of phase-pure $La_{n+1}Ni_nO_{3n+1}$ thin films with $n > 3$ has been reported in the literature, likely due to thermodynamic instability. To grow the film by ALL-Laser MBE, $La_2O_3$ and NiO targets were ablated alternately to deposit the $NiO_6$ octahedron layers four times before an additional LaO layer was added. The RHEED intensity oscillation, shown in Fig. 3B, contains twice as many laser pulses in the fourth LaO deposition in each period as the other LaO layers. The XRD $\theta$-$2\theta$ scan of the resultant 50 unit-cell La$_5$Ni$_4$O$_{13}$ film is presented in Fig. 3C, showing the diffraction peaks of the $n = 4$ RP phase from 00$\underline{10}$ to 00$\underline{28}$ without any impurity peaks or peak splitting. Figure 3D is a cross-sectional scanning transmission electron microscope (STEM) high angle annular dark field (HAADF) image of the La$_5$Ni$_4$O$_{13}$ film, showing an extra layer of LaO every 4 unit cells of LaNiO$_3$. In Fig. 3E, atomic resolution elemental mapping using electron energy-loss spectroscopy (EELS) indicates a perfect match between the elemental distribution and the simultaneously taken HAADF image, verifying that the extra layer is indeed LaO. Evidently, ablating from the oxide targets of La$_2$O$_3$ and NiO is able to produce the polar LaO and NiO$_2$ layers to form the Ruddlesden-Popper structure. A change in the Ni valence next to the LaO rock salt layer must take place to maintain the charge neutrality in the Ruddlesden-Popper phase (*18*).

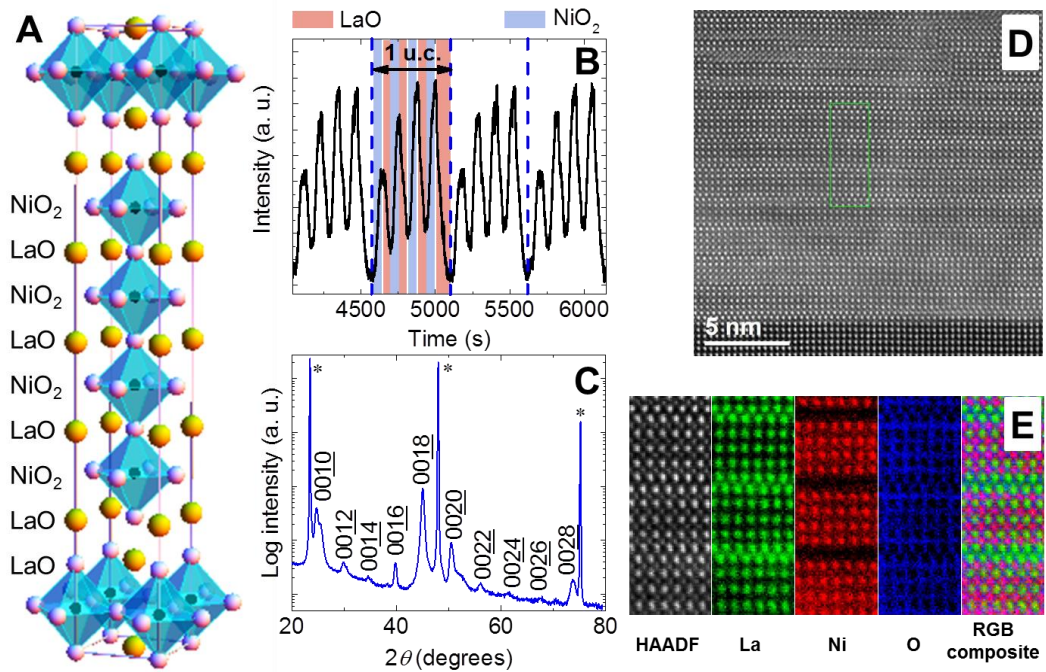

**Fig. 3.** (**A**) Schematic of the La$_{n+1}$Ni$_n$O$_{3n+1}$ structure with $n = 4$, La$_5$Ni$_4$O$_{13}$. (**B**) RHEED intensity oscillations during the growth of a La$_5$Ni$_4$O$_{13}$ film. An extra LaO layer is inserted

after every four pairs of NiO$_2$ - LaO layers are deposited. (**C**) XRD $\theta$-$2\theta$ scan of the La$_5$Ni$_4$O$_{13}$ film. (**D**) STEM HAADF image of the La$_5$Ni$_4$O$_{13}$ film. (**E**) Simultaneously taken HAADF image and EELS elemental mapping for the area marked with a rectangular box in Fig. 3D.

Since the discovery of the 2D electron gas at the LaAlO$_3$/SrTiO$_3$ interface (*2*), several competing mechanisms for its origin have been proposed and intensely debated, including electronic reconstruction (*19*), oxygen vacancies in the SrTiO$_3$ substrate (*20, 21, 22*), and intermixing between the LaAlO$_3$ film and the SrTiO$_3$ substrate (*23, 24*). According to the electronic reconstruction mechanism, because the atomic layers are charge neutral in SrTiO$_3$ but charged in LaAlO$_3$, a diverging electric potential is built up in a LaAlO$_3$ film grown on a TiO$_2$-terminated SrTiO$_3$ substrate. This leads to the transfer of half of an electron from the LaAlO$_3$ film surface to SrTiO$_3$ when the LaAlO$_3$ layer is thicker than 4 unit cells, creating a 2D electron gas at the interface with a sheet carrier density of $3.3 \times 10^{14}$/cm$^2$ when LaAlO$_3$ is sufficiently thick. A serious inconsistency with this mechanism is that the carrier densities reported experimentally are invariably lower than the expected value (*25, 26*) except under conditions where reduction of SrTiO$_3$ substrate is suspected (*20, 21*). Oxygen vacancies in SrTiO$_3$ are known to contribute to conductivity, but all reported conducting LaAlO$_3$/SrTiO$_3$ interfaces have been grown at low oxygen pressures (< 10 mTorr), and annealing in oxygen is often required (*1, 19, 27, 28*); higher oxygen pressures during the PLD growth result in insulating samples (*27*) or 3D island growth (*29*). Low growth pressures can not only cause oxygen vacancies in SrTiO$_3$, but can also enhance the bombardment effect due to energetic species that may lead to La-Sr intermixing at the interface (*27*). At present, there is no consensus on the origin of the 2D electron gas at the LaAlO$_3$/SrTiO$_3$ interface.

With ALL-Laser MBE, we grew LaAlO$_3$ film one atomic layer at a time, which allowed us to produce conducting LaAlO$_3$/SrTiO$_3$ interfaces at an oxygen pressure as high as 37 mTorr. This high oxygen pressure helps to prevent oxygen reduction in SrTiO$_3$, ensure that the LaAlO$_3$ films are sufficiently oxygenated, and suppress the La-Sr intermixing due to the bombardment effect. Furthermore, we grew LaAlO$_3$ films of different cation stoichiometry, LaAl$_{1+y}$O$_{3(1+0.5y)}$, as a way to test the electronic reconstruction hypothesis. As depicted by Sato *et al*. (*28*), either Al vacancies or La vacancies in the off-stoichiometric films lead to oxygen vacancies in order to keep the charge neutrality. As a result, the charges on each layer depend on the stoichiometry: instead of +1 for the [LaO] layer and -1 for the [AlO$_2$] layer, the charge is $+(1-2y)$ on the [LaO$_{1+y}$] layer and $-(1-2y)$ on the [Al$_{1+y}$O$_{2+0.5y}$] layer. A similar modification of charges in each layer has been suggested for LaAlO$_3$ films diluted with SrTiO$_3$ (*30*). In the electronic reconstruction picture, instead of the charge transfer of 0.5$e$ in the case of stoichiometric LaAlO$_3$, $(0.5-y)e$ will be transferred to resolve the polar discontinuity at the interface. The sheet carrier density depends linearly on $y$, *i.e.* $n_s = (1-2y) \times 3.3 \times 10^{14}$/cm$^2$ for sufficiently thick LaAlO$_3$ (*28*). For the 10 unit-cell LaAlO$_3$ films used in this work, the dependence becomes $n_s = (1-2y) \times 3.3 \times 10^{14}$/cm$^2 - 1.6 \times 10^{14}$/cm$^2 = (1.7-2y \times 3.3) \times 10^{14}$/cm$^2$ (*31*).

Starting from a TiO$_2$-terminated SrTiO$_3$ substrate, we grew 10 unit-cell-thick LaAlO$_3$ films by alternately ablating La$_2$O$_3$ and Al$_2$O$_3$ targets under an oxygen pressure of 37 mTorr. The RHEED intensity oscillations during the growth of a stoichiometric film are shown in Fig. 4A. The RHEED intensity of the first LaO layer shows an irregular pattern, reflecting the state of surface chemistry, charge, and morphology during the transition from TiO$_2$ to LaO. The pattern becomes more regular in the subsequent layers. The reduction in the RHEED intensity from the substrate level is small throughout the growth of the 10 unit-cell stoichiometric LaAlO$_3$ film. The 2D growth mode was maintained as confirmed by the sharp

RHEED spots in Fig. 4B for the 10 unit-cell film. The atomic force microscopy (AFM) image for the film in Fig. 4C shows an atomically flat surface. To change the cation stoichiometry of the film, the number of laser pulses on the $Al_2O_3$ target for each $Al_{1+y}O_2$ layer was varied.

Because of the high oxygen pressure during the $LaAlO_3$ growth, the samples were well oxygenated. This was proven by polarization-dependent x-ray absorption spectroscopy (XAS) measurements. Figure 4D shows XAS spectra with different linear polarizations for a stoichiometric $LaAlO_3$ film and Fig. 4E shows the Ti $L_{2,3}$ x-ray linear dichroism (XLD) signals for different $LaAl_{1+y}O_{3(1+0.5y)}$ stoichiometry. No $Ti^{3+}$ related features around 462 eV, characteristic of the oxygen deficient $LaAlO_3/SrTiO_3$ samples (*32, 33*), are observed. Rather, the spectra are similar to the fully oxygenated samples (*32, 33*). Figure 4F shows the Ti $L_{2,3}$ x-ray magnetic circular dichroism (XMCD) signals obtained from opposite circularly polarized XAS spectra. Very small XMCD signals were observed, indicating very weak ferromagnetism. This again is consistent with the fully oxygen annealed $LaAlO_3/SrTiO_3$ samples (*32, 33*).

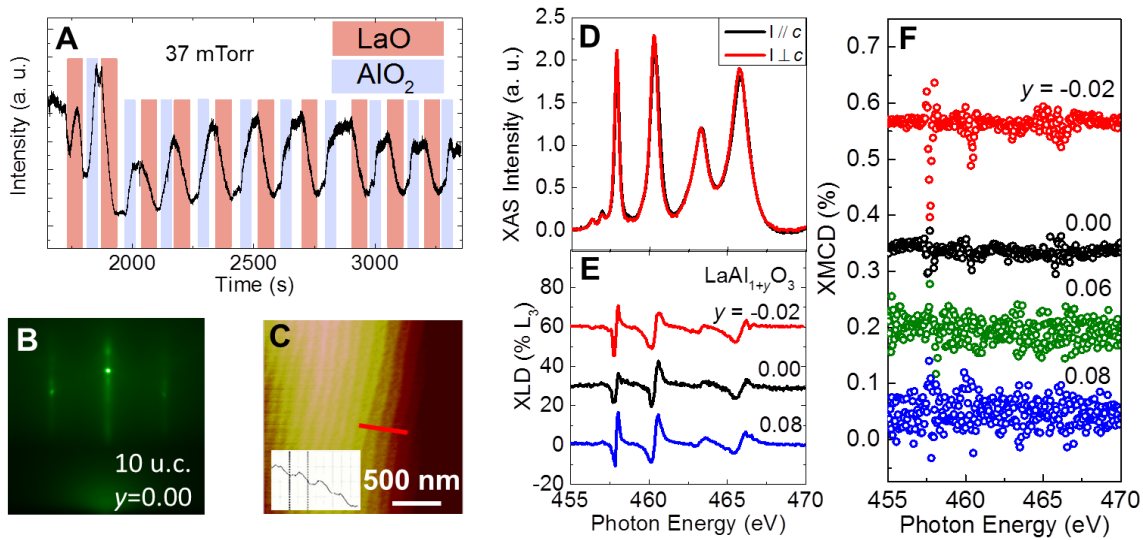

**Fig. 4.** (**A**) RHEED intensity oscillation during the growth of a $LaAlO_3$ film on a $SrTiO_3$ substrate at an oxygen pressure of 37 mTorr. (**B**) RHEED pattern after the growth of a 10

unit-cell stoichiometric LaAlO$_3$ film. (**C**) AFM topographic images of the 10 unit-cell LaAlO$_3$ film on SrTiO$_3$. The root mean square (RMS) roughness of the film is 0.1 nm. (**D**) Ti XAS spectra measured at 13 K with a polarization parallel (black curve) and perpendicular (red curve) to the sample normal for a 10 unit-cell stoichiometric LaAlO$_3$ film on SrTiO$_3$. (**E**) Ti $L_{2,3}$-edges XLD spectra for 3 LaAl$_{1+y}$O$_{3(1+0.5y)}$/SrTiO$_3$ samples with $y$ = -0.02, 0.00, and 0.08. (**F**) XMCD observed for the in-plane geometry for 4 LaAl$_{1+y}$O$_{3(1+0.5y)}$ films on SrTiO$_3$ with $y$ = -0.02, 0.00, 0.06, and 0.08. All samples were measured at a temperature of 13 K in a constant field of ±0.3 T.

The temperature and stoichiometry dependences of the sheet resistance, sheet carrier density, and mobility are shown in Fig. 5A-F, respectively, for the 10 unit-cell thick LaAl$_{1+y}$O$_{3(1+0.5y)}$ films. All of the films are conducting with sheet resistance around $10^4$ Ω/□ at 300 K, in contrast to insulating films grown by PLD from LaAlO$_3$ compound targets at this oxygen pressure (*27, 34*). Note that only the Al-rich LaAl$_{1.08}$O$_{3.12}$ film shows metallic behavior in the full temperature range, while all other films show low-temperature resistivity upturns, consistent with the previous stoichiometry dependence reports (*35, 36*). The low temperature upturn has a -ln$T$ dependence characteristic of the Kondo effect (*37*). This may be attributed to the inevitable defects at the LaAlO$_3$/SrTiO$_3$ interface, consistent with the weak magnetism shown by Fig. 4F. The black dashed line in Fig. 5D represents the threshold normal-state sheet resistance $h/4e^2$, or 6.5 kΩ/□ for a superconductor-insulator transition (*38*). Only the Al-rich samples have normal-state sheet resistance below the dashed line; these might thus exhibit superconductivity. The sheet carrier density is around $10^{14}$/cm$^2$ for all the samples, close to the expected value of $1.7 \times 10^{14}$/cm$^2$. The sample with a higher sheet carrier density shows a lower mobility, in agreement with previous reports (*26*).

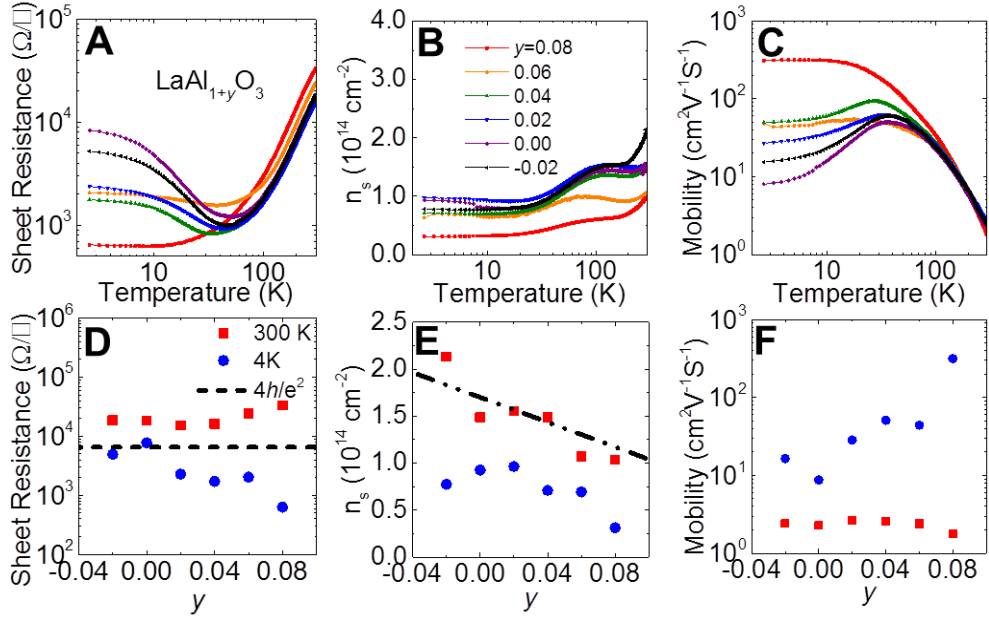

**Fig. 5.** (**A**) Sheet resistance, (**B**) carrier density, and (**C**) Hall mobility as functions of temperature for a series of 10 unit-cell $LaAl_{1+y}O_{3(1+0.5y)}$ films with different $y$ values. (**D**) Sheet resistance, (**E**) sheet carrier density, and (**F**) Hall mobility as functions of film stoichiometry at 300 K (red squares) and 4 K (blue dots). The dashed line in (**D**) is the quantum resistance limit $h/4e^2$ (*38*). The dashed line in (**E**) indicates the theoretical value of sheet carrier density for 10 unit-cell films with different stoichiometry under the assumption of pure electronic reconstruction (*31*).

The central result of this work is the dependence of sheet carrier density on the stoichiometry of $LaAl_{1+y}O_{3(1+0.5y)}$ shown in Fig. 5E. The black dashed line represents $n_s = (1.7-2y \times 3.3) \times 10^{14}/cm^2$, which is expected by the electronic reconstruction hypothesis (*31*). The red squares, which denote 300 K sheet carrier density, overlap with the dashed line. No additional mechanism is employed to explain our data, and we cannot produce the linear dependence on the $LaAl_{1+y}O_{3(1+0.5y)}$ stoichiometry using either oxygen vacancies or intermixing mechanisms without invoking remarkable coincidence. The quantitative agreement between our experimental result and the theoretical prediction provides a strong support to the electronic reconstruction mechanism as responsible for the 2D electron gas at the LaAlO$_3$/SrTiO$_3$ interface.

The key differences between our result and the previous reports are the high oxygen pressure during the film growth and the high film crystallinity as demonstrated by Fig. 4. We argue that the high oxygen pressure suppresses the likelihood of oxygen vacancies in SrTiO$_3$ and the bombardment effect. The *c* lattice constant of a stoichiometric LaAlO$_3$ film in this work was small (3.71 Å), consistent with the expected value based on the Poisson ratio for a coherently strained film on SrTiO$_3$, consistent with it having high structural perfection. Previous PLD studies on the stoichiometry dependence varied the laser energy density and oxygen pressure to change the LaAlO$_3$ composition (*28, 36*). In our study the films were grown under the same conditions, therefore, the stoichiometry dependence observed is free from the effects of varying deposition conditions.

**Acknowledgments:**

We thank Dr. S. L. Shi and Dr. F. Q. Huang of Shanghai Institute of Ceramics, Chinese Academy of Sciences for synthesizing the ceramic SrO target which was used when developing the ALL-laser MBE technique. We thank Dr. P. S. Risborough for the helpful discussions concerning the Kondo effect. We thank Dr. Ph. Ghosez for comments on the manuscript and providing us with the first-principles calculation data on the LaAlO$_3$ thickness dependence.

This material is based upon work supported by the U.S. Department of Energy, Office of Science, under Grant No. DE-SC0004764 (Q. Y. L. and X. X. X.). Raman studies at Boise State University have been supported by NSF under Grant No. DMR-1006136 (A.K.F. and D. A. T.) TEM study was supported by CCDM, an EFRC funded by U.S. DOE-BES, under award #DE-SC0012575 (Q.Q.), by DOE-BES, Materials Science and Engineering, under Contract No. DE-SC0012704 (Y.Z.), and used resources of CFN at BNL, a U.S. DOE Office of Science Facility. A.X.G. acknowledges support from the U.S. Army Research Office, under Grant No. W911NF-15-1-0181. The Advanced Light Source is supported by the Director, Office of Science, Office of Basic Energy Sciences, of the U.S. Department of Energy under Contract No. DE-AC02-05CH11231.

**Supplementary Materials:**

Materials and Methods

Figures S1-S6

Reference (*39-48*)

**Materials and Methods**

<u>Sample preparation and characterization</u>

The (001) SrTiO$_3$ substrates used in this work were treated following the receipe in (*39*) to produce atomically flat TiO$_2$-terminated surface with one unit-cell-high steps (Fig. S1A). The (001) LaAlO$_3$ substrates used in this work were treated following the receipe in (*40*) to produce atomically flat AlO$_2$-terminated surface with one unit-cell-high steps (Fig. S1B). The surface morphology of a 60 nm SrTiO$_3$ (Fig. S1C) grown on treated SrTiO$_3$ substrates also shows atomically flat surface with root mean square (RMS) roughness around 0.1 nm, comparable to that of the SrTiO$_3$ substrate.

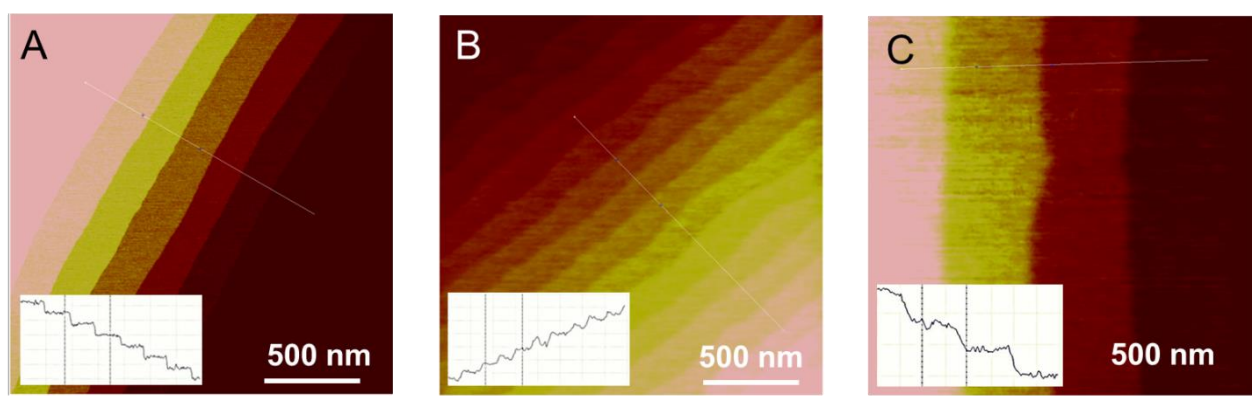

**Fig. S1.** AFM topographic images for (**A**) TiO$_2$ terminated SrTiO$_3$ (001) surface, (**B**) AlO$_2$ terminated LaAlO$_3$ (001) surface, and (**C**) 60 nm stoichiometric SrTiO$_3$ film grown on SrTiO$_3$ substrate. The inserts show the cross section of the image along the steps, indicating

that atomically flat surfaces with step-and-terrace structures have been obtained for the substrates and the film. RMS roughness for all the substrates and the film are around 0.1 nm.

A KrF excimer laser ($\lambda$ = 248 nm, pulse duration 25 ns) was used with repetition rate ranging from 1 Hz to 30 Hz. For the growth of SrTiO$_3$, a single crystal SrO target and a ceramic TiO$_2$ target were used. The oxygen pressure during the growth was $1\times10^{-6}$ Torr and the substrate temperature was 760 °C. With a laser energy density of 0.9 J/cm$^2$, the number of laser pulses was about 100 for each SrO layer and 80 for each TiO$_2$ layer in the growth of stoichiometric SrTiO$_3$ film. The offstoichiometric films Sr$_{1+x}$Ti$_{1-x}$O$_3$ were created by supplying 100$x$ % more SrO and TiO$_2$ in each layer. After the deposition, the films were cooled to room temperature in oxygen at the same pressure as during the growth. For the growth of La$_5$Ni$_4$O$_{13}$, ceramic targets of La$_2$O$_3$ and NiO were used. The oxygen pressure during the growth was 37 mTorr and the substrate temperature was 600 °C. With the laser energy densities of 0.7 J/cm$^2$ for LaO and 0.9 J/cm$^2$ for NiO, the number of laser pulses was 75 for each LaO layer and 107 for each NiO$_2$ layer. The sequence for the growth of each Ruddlesden Popper unit cell is LaO-NiO$_2$-LaO-NiO$_2$-LaO-NiO$_2$-LaO-NiO$_2$-LaO, *i. e.* four LaNiO$_3$ layers followed by one extra LaO layer. After the deposition, the films were cooled to room temperature at a oxygen pressure of $8 \times 10^4$ Pa. For the growth of LaAlO$_3$ film, ceramic targets of La$_2$O$_3$ and Al$_2$O$_3$ were used. The oxygen pressure during the growth was 37 mTorr and the substrate temperature was 720 °C. The laser energy density was 1.0 J/cm$^2$ on the La$_2$O$_3$ target, requiring 75 laser pulses for each LaO layer, and 1.3 J/cm$^2$ on the Al$_2$O$_3$ target, requiring 91 laser pulses for each AlO$_2$ layer. The offstoichiometric LaAl$_{1+y}$O$_{3(1+0.5y)}$ films were created by supplying 100$y$ % more Al in each layer. After the deposition, the films were cooled to room temperature in oxygen at the same pressure as during the growth.

The x-ray diffraction and reflectivity measurements were performed using a Bruker D8 Discover system with Cu K$_\alpha$ radiation ($\lambda$ = 1.5406 Å). The Leptos fitting software (Bruker AXS Inc.) was used to determine the out-of-plane lattice constant of the films from the $\theta$-$2\theta$ scans (the substrate peak was used as the reference) and to determine the film thickness from the x-ray reflectivity (XRR) measurement.

Ultraviolet Raman spectroscopy measurements were performed in a backscattering geometry normal to the film surface using a Jobin Yvon T64000 triple spectrometer equipped with a liquid nitrogen cooled multichannel charge coupled device detector. Ultraviolet light (325 nm line of the He-Cd laser) was used for excitation. Maximum laser power density was ~0.5 W/mm2 at the sample surface, low enough to avoid any noticeable local heating of the sample. Spectra were recorded at 10 K using a variable temperature closed-cycle helium cryostat.

X-ray absorption spectroscopy (XAS), x-ray linear dichroism (XLD) and x-ray magnetic circular dichroism (XMCD) measurements were carried out at the at the elliptically polarized undulator beamline 4.0.2 of the Advanced Light Source, using the Vector Magnet end station and with an energy resolution of approximately 0.1 eV (*41*). Samples were cryogenically cooled to 13 K. Average probing depth in the total electron yield XAS detection mode was estimated to be approximately 5 nm, providing interface-sensitive information with minimal contribution from surface adsorbates. Measurements were carried out in near-grazing (30°) incidence geometry, enabling selective alignment of the x-ray electric field parallel to the *ab*-plane of the film for vertically-polarized light ($E \parallel ab$), and almost parallel to the *c*-axis of the film for vertically polarized light ($E \parallel c$). The XMCD spectra were obtained utilizing circularly polarized x-rays with and by alternating the direction of the applied magnetic field of 0.3 T between parallel and antiparallel directions with respect to the x-ray helicity vector.

RHEED intensity oscillation patterns of the specular spot, the diffraction spots, and the Kikuchi line intersection

For atomic layer-by-layer film growth by reactive MBE of ALL-laser MBE, the intensity of the RHEED diffraction spot along the $SrTiO_3$ [110] azimuth rather than the specular spot is monitored and used to control the growth. The specular spot intensity often shows "double peak" behavior and is less predictable. This effect has been recognized and articulated by Dobson *et al*. as due to the superposition of the elastic specular scattered and the diffused scattering such as Kikuchi bands (*42*). Figures S2A shows the RHEED pattern after the deposition of a $TiO_2$ layer. One can clearly see that beside the specular spot and the diffraction spot, there is a bright spot due to the intersection of Kikuchi lines. Figure S2B shows the RHEED intensity oscillations corresponding to different areas of integration as marked by the various rectangles as the targets of SrO and $TiO_2$ were being alternately ablated. The intensity of the Kikuchi intersection spot oscillates in-phase with the diffraction spot with the same period. The specular spot also oscillates with the same period as the diffraction spot, but 180° out of phase. When the Kikuchi spot and the specular spot are integrated together, a complex intensity oscillation pattern appears. In principle, the specular spot intensity oscillation can be used for the control of atomic layer-by-layer growth; in practice, however, it often overlaps with the Kikuchi spot, making its use difficult.

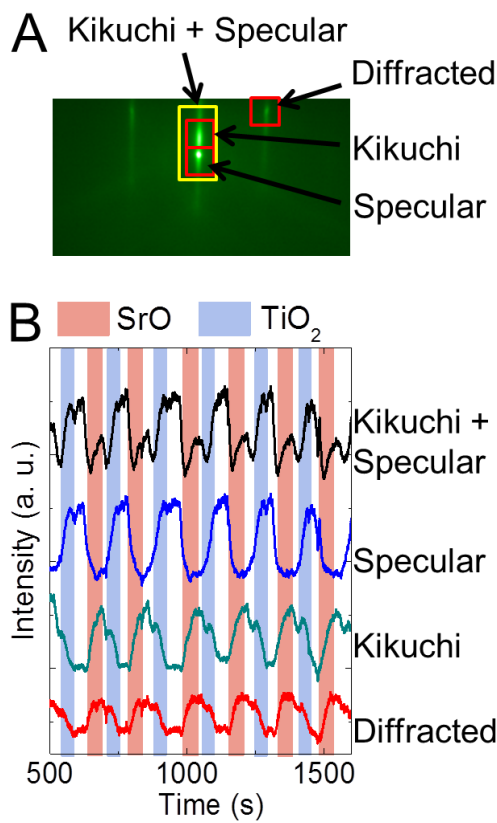

**Fig. S2.** (**A**) RHEED pattern after the growth of a TiO$_2$ layer. (**B**) RHEED intensity oscillations for specular spot, diffraction spots, and Kikuchi line intersection.

Growth calibration using split RHEED intensity peaks

Strontium, with a larger atomic number, has a larger scattering power, or atomic form factor, than Ti (*43*). Therefore, in general, the deposition of the SrO layer leads to an increase in the RHEED intensity and the growth of the TiO$_2$ layer causes the RHEED intensity to decrease. Surface roughness adds complexity to this simple picture and the combined effects of roughness and chemistry give rise to the experimentally observed variety of RHEED intensity oscillation patterns. As shown in Fig. 1B in the main text, an oversupply of Sr causes the RHEED intensity peak to split. By adding one half extra SrO layer on top of stoichiometric SrTiO$_3$, split RHEED intensity peaks as those in Fig. S3A can be created. Four films are grown with this double peak calibration and the growth for each film is interrupted at different time as shown in Fig. S3B. Starting at time $t_1$, a film with half monolayer (ML) of SrO on top of a completed TiO$_2$ ML is presented in Fig. S3C. The surface roughness R$_q$ of this film is 0.13 nm, slightly larger than that of a TiO$_2$ terminated substrate (0.11 nm), indicating an incomplete coverage. RHEED intensity for this film is as indicated at $t_1$ in Fig. S3B. As the deposition of SrO continues to $t_2$, where 1.2 ML of SrO was deposited, RHEED intensity increases to a maximum. Rq of this film drops to 0.12 nm, indicating a better coverage of the surface. The growth for the third film continued SrO deposition to time $t_3$, where 1.5 ML SrO was completed. The surface roughness increase to 0.17 nm, indicating some extra rock-salt structure of SrO on the surface as shown in Fig. S3E. From $t_3$ to $t_4$, the target is switched from SrO to TiO$_2$. For the forth film, where the growth is stopped after 0.4 ML TiO$_2$ is deposited on top of 1.5 ML SrO, RHEED reaches its local maxima. Smoothing happens from $t_4$ to $t_5$, as the Rq drops from 0.17 nm to 0.13 nm. Some segregation between SrO and TiO$_2$ (*44*) may take place during this period of time as well, indicated by a local minima between $t_4$ and $t_5$. From $t_5$ to $t_6$, a ML of TiO$_2$ is completed. The segregation continues and the surface structure becomes to 0.5 ML SrO on top of 1 ML TiO$_2$, which is the same as $t_1$. Thus, RHEED intensity for $t_1$ and $t_6$ are identical. A complete ML of SrO and TiO$_2$ are deposited during t1 to t6, and RHEED oscillation finishes a whole period. As long as a stoichiometric supply of SrO and TiO$_2$ is maintained, the split RHEED intensity pattern will remain. This effect is utilized to more accurately calibrate the numbers of laser pulses for each SrO or TiO$_2$ layer.

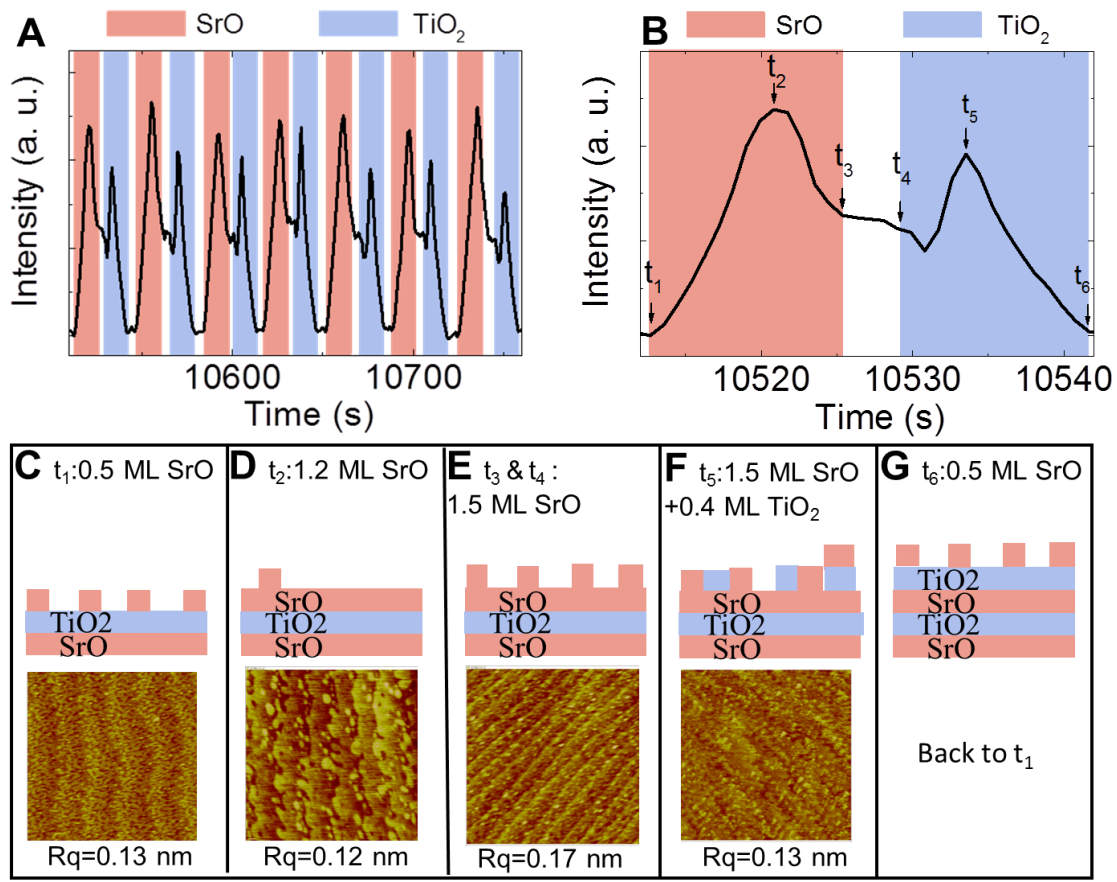

**Fig. S3.** (**A**) RHEED intensity oscillations for a stoichiometric SrTiO$_3$ growth on top of one half extra SrO layer. (**B**) A zoom in on one of the RHEED oscillation in Fig. S3A. (**C**)-(**G**) Schematics, AFM images and surface root mean square roughness (Rq) for five films with growth stopped at certain time from $t_1$ to $t_6$ in Fig S3B.

Deposition rate calibration by thickness measurement

Besides the beating in RHEED intensity oscillation shown in Fig. 1C, the thickness measurement of calibration films by XRR was also used to calibrate the SrO and TiO$_2$ deposition rates for complete monolayer coverages. Because XRR oscillations are usually not observable for a stoichiometric SrTiO$_3$ film on SrTiO$_3$ substrate (*45*), LaAlO$_3$ substrate was used for the calibration films. Figure S4 shows the XRR scan and fitting curve for a SrTiO$_3$ film built with 154 pairs of SrO and TiO$_2$ layers. The thickness of the film from the XRR measurement is 62.7 nm as compared to the thickness of 154 unit cells fully relaxed SrTiO$_3$ 60.1 nm. Assuming that the SrTiO$_3$ film on LaAlO$_3$ substrate is fully strained and considering the Poisson's ratio $\upsilon$=0.23 (*46*) for SrTiO$_3$, the measured thickness agrees with the designed thickness within 2% (*47*).

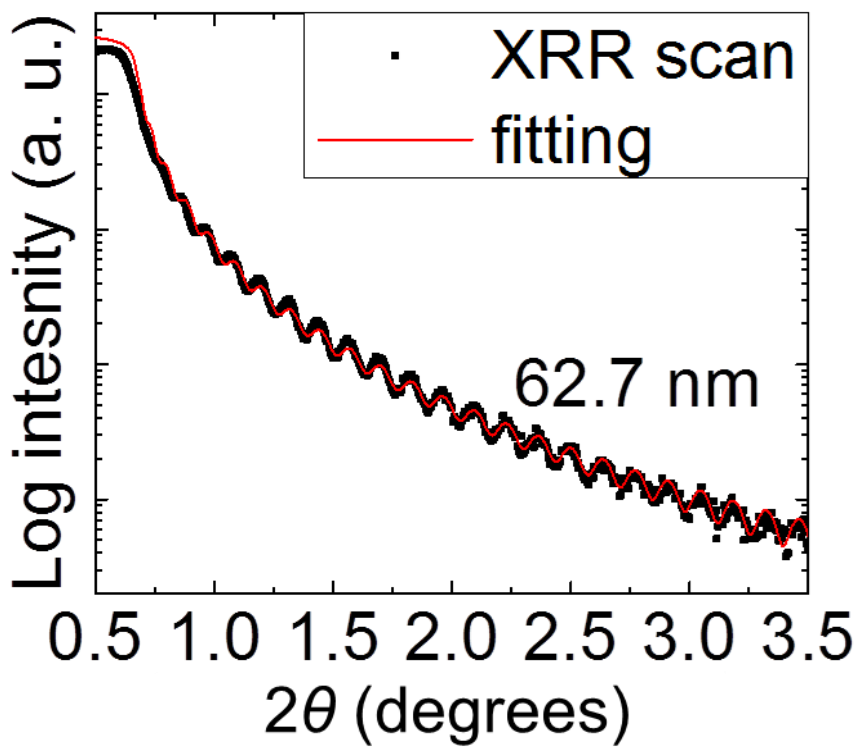

**Fig. S4.** XRR measurement and fitting for 154 unit cells of stoichiometric SrTiO$_3$ on LaAlO$_3$ substrate.

Structure of a 25 u.c. LaAlO$_3$ film on SrTiO$_3$ substrate.

The 25 u.c. LaAlO$_3$ film was grown from La$_2$O$_3$ and Al$_2$O$_3$ targets under the calibrated stoichiometric growth condition in an oxygen pressure of 37 mTorr. The film *c* lattice constant was determined to be 3.71 Å by XRD $\theta$-$2\theta$ scan and fitted with Nelson-Riley method. Using a Poisson ratio $\upsilon$ =0.26 (*48*), and LaAlO$_3$ bulk *c* lattice constant 3.79 Å, the *c* lattice constant calculated for the film is 3.71 Å.

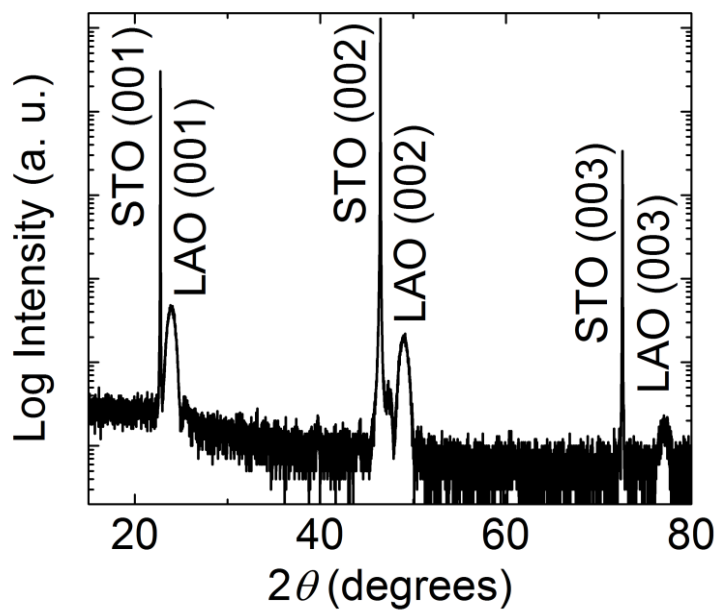

**Fig. S5.** XRD $\theta$-$2\theta$ scan of a 25 u.c. LaAlO$_3$ film grown on SrTiO$_3$ substrate.

STEM image of a stoichiometric SrTiO$_3$ film on SrTiO$_3$ substrate

Figure S6 is a scanning transmission electron microscope (STEM) cross-sectional image of the stoichiometric film. A small contrast along the interface was observed. The SrTiO$_3$ film is free of off-stoichiometry related defects, making it indistinguishable from the SrTiO$_3$ substrate. No Ruddlesden–Popper planar faults disordered structure were observed, indicating film stoichiometry.

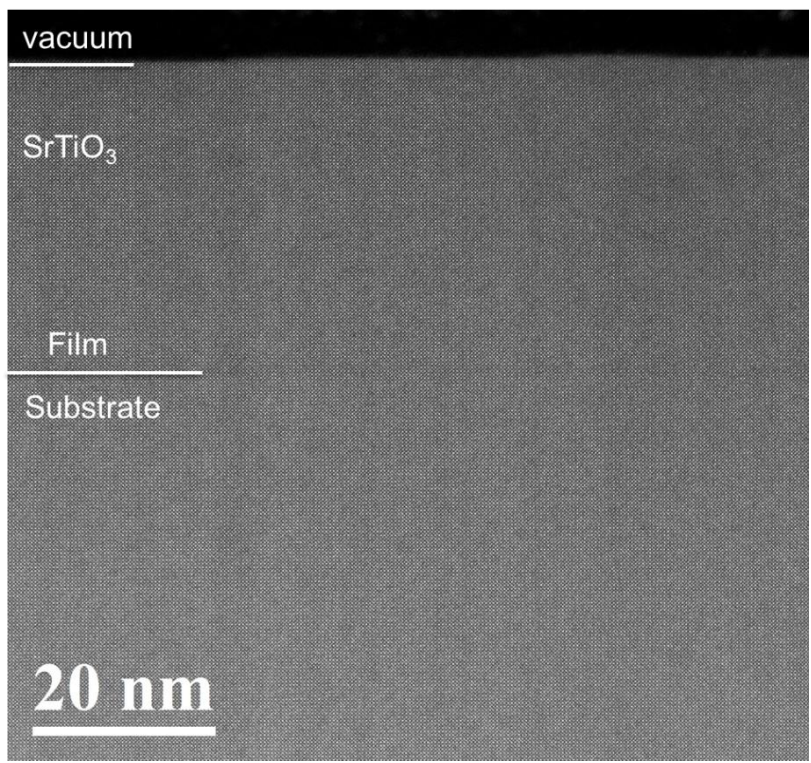

**Fig. S6.** Cross-sectional STEM HAADF image of a stoichiometric 30 nm film.